# On the Existence of Photoluminescence and Room-Temperature Spin Polarization in Ambipolar V doped $MoS_2$ Monolayers

Dipak Maity[1], Rahul Sharma[1,*], Krishna Rani Sahoo[1], Ashique Lal[1], Raul Arenal[2,3,4], Tharangattu N. Narayanan[1,*]

[1]Tata Institute of Fundamental Research – Hyderabad
500046, INDIA.

[2]Instituto de Nanociencia y Materiales de Aragon (INMA), Universidad de Zaragoza
50009 Zaragoza, SPAIN.

[3]Laboratorio de Microscopias Avanzadas (LMA), Universidad de Zaragoza, 50018 Zaragoza, SPAIN.

[4]Fundación ARAID, 50018 Zaragoza, SPAIN.

(*corresponding Authors: tnn@tifrh.res.in, rahulsharmatifr@gmail.com)

**Abstract:**

Opto-spintronics is an emerging field where ultra-thin magnetic-semiconductors having high spin-valley coupling play an important role. Here, we demonstrate substitutional vanadium (V) doping in $MoS_2$ lattice in different extent, leading to the coexistence of photoluminescence (PL), valleypolarization (~32%), and valley splitting (~28 meV shift in PL with helicity $\sigma^+$ and $\sigma^-$ of light excitation). A large V doping causes semiconductor to metal transition in $MoS_2$ but with medium level causing the existence of photoluminescence with high spin polarization. The ambipolar nature of medium level V doped $MoS_2$ is shown here indicating its potential as an opto-electronic material. The presence of V-dopants and their different level of content are proven by both spectroscopic and microscopic methods. A detailed temperature and power dependent photoluminescence studies along with density functional theory-based calculations in support unravels the emergence of the co-existence of spin-valley coupling and photoluminescence. This study shows the potential of doping $MoS_2$ for deriving new materials for next generation room temperature opto-spintronics.

## 1. Introduction:

Developing magnetically ordered semiconductors is highly demanding for next generation spintronic devices[1–3], while many of the existing magnetic materials are metallic in nature.[4] Recent advancements in atomic layers, particularly in a class of materials called transition metal dichalcogenides (TMDs), and their implications in ultra-thin optoelectronic devices are leading to the development of opto-electronic devices based on this new class of materials[5–8]. Though the heavy metals based TMDs have inherent high spin orbit coupling (SOC), they do not exhibit magnetic ordering, particularly at room temperature.[3,9–11] Molybdenum disulfide ($MoS_2$) is one such important TMD explored largely, but lacking any ferromagnetic ordering at room temperature.[12,13] Similar to graphene[14–16], engineering the defects or with other heavier elements or ferromagnetic impurity atoms are shown to be leading to magnetic ordering in monolayer $MoS_2$, though the development of room temperature ferromagnetic semiconducting $MoS_2$ was far from the reach.[17–19]

$MoS_2$, particularly that grown *via* chemical vapor deposition (CVD), is an n-type material (from the transport studies[20–23]). The presence of sulfur vacancy can be attributed to this n-type behavior of $MoS_2$ though the ultra-violet photoelectron spectroscopic studies did not reveal this n-type nature.[24,25] But it has been shown theoretically and experimentally that sulfur vacancy provides deep acceptor levels near the conduction band and hence it cannot cause n-type conductivity[24–26], instead trapped charges at the substrate-$MoS_2$ interface can be responsible for such n-type nature. Our previous studies on $MoS_2$ (on $SiO_2$/Si substrate) field effect transistor (FET) based transfer studies also revealed the n-type nature of CVD grown pristine $MoS_2$, while interfacing with fluorographene or doping can make it p-type or ambipolar in nature.[5,27–29] The presence of Schottky barrier at the metal electrode and $MoS_2$ interface can increase the contact resistance and, hence adversely affect the electrical performance of the $MoS_2$ based devices. Furthermore, presence of interfacial states and traps in $MoS_2$ is also found to be leading to Fermi level pinning while making contacts with metal current collectors.[30–32] Ambipolar semiconducting $MoS_2$ layers can perform better in opto-electronic devices, since the presence of transport of both holes and electrons in $MoS_2$ indicates that either the Fermi-level pinning position has been changed in the $MoS_2$ band gap or it has been unpinned.[29]

Recently, vanadium (V) doping in TMDs has shown to be bringing interesting attributes to these atomic layers such as valley polarization, tunable carrier density, dilute magnetism etc.[33–37] It has been shown that valence band splitting due to V doping is also high due to its high g factor.[37] Additionally, theoretical studies predict that V-doping can also introduce spin polarization in MoS$_2$ band structure.[38] Despite a few works on V doped TMDs of WSe$_2$ and MoS$_2$, a detailed study on the photoluminescence, electrical, and magnetic properties of V doped MoS$_2$ (VMS) is still lacking, where such a study can unravel the nature of energy states forming due to vanadium. Here we study both theoretically and experimentally the changes in the optical, electronic, and magnetic properties of monolayer MoS$_2$ (MS) with different amounts of V doping.

VMS monolayers are synthesized using a CVD technique and have been characterized by transmission electron microscopy (TEM) and X-ray photoelectron spectroscopy (XPS). The transport properties of V-doped MS areconducted to understand the p doping of V atoms with varying V concentration. Effects of doping on the work function and carrier density are also studied using transport measurements. Low-temperature photoluminescence (PL) study of the VMS reveals the formation of a new excitonic state (V-peak) at ~ 850 nm. A detailed temperature and power-dependent PL study is conducted to understand the origin of this new state. Experimental observations are further verified using density functional theory (DFT) based first principal calculations. The valley-polarization in VMS is verified using polarization selective PL experiments explained in the later part.

## 2. Result and discussion:

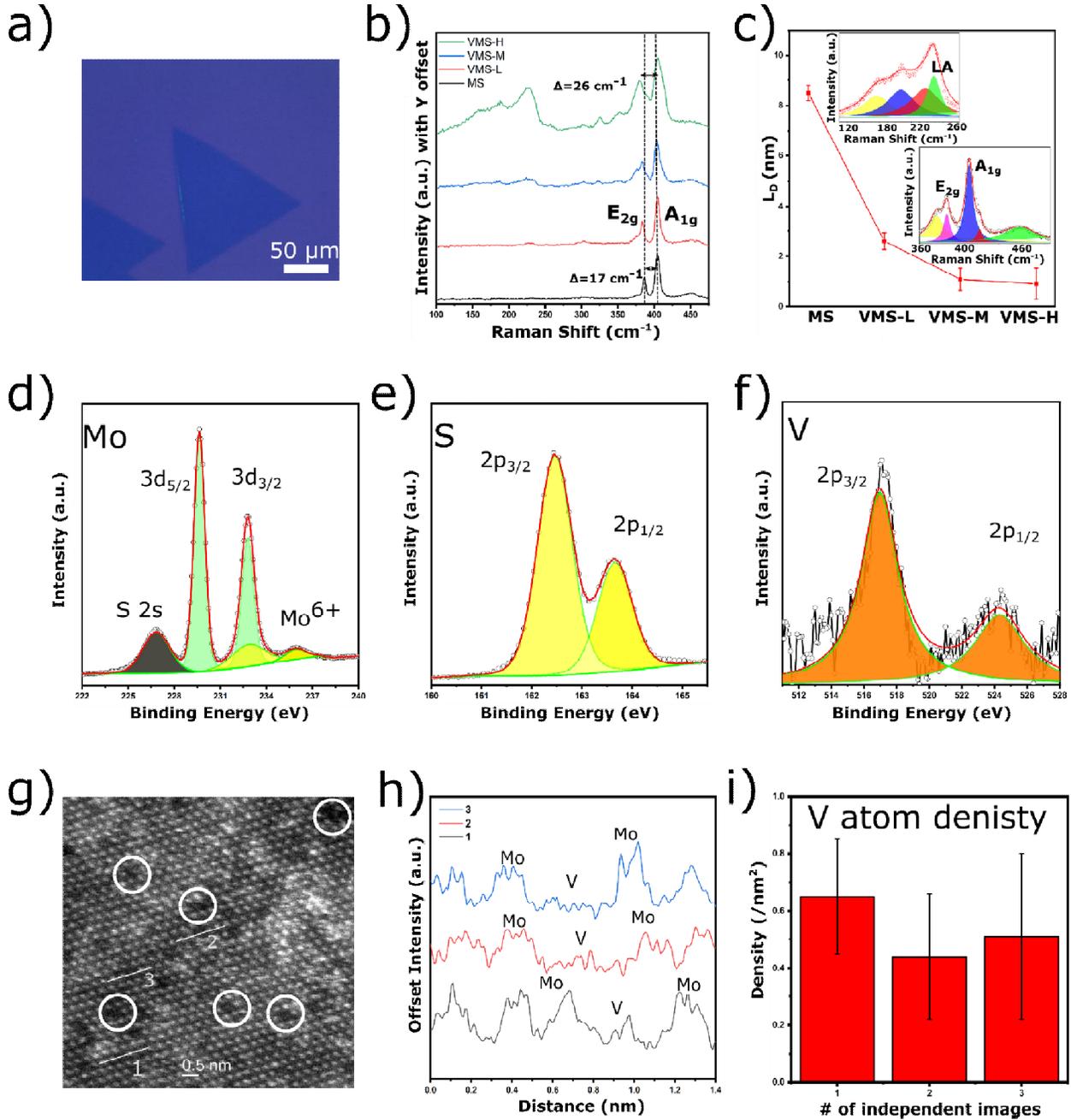

*Figure 1: a) Optical image of a monolayer VMS. VMS crystals of more than 100 μm width are shown to be formed. b) Raman spectra of MS and VMS having three different levels of V contents. c) $L_D$ as a function of different level of V contents. High resolution XPS spectra of VMS-M for d) Mo, e) S, and f) V. g) STEM-HAADF image of VMS monolayer showing the*

*presence of V dopants. h) Line profiles extracted from the STEM image of Figure 1g, see highlighted lines in the micrograph.i)V atom density in three independent STEM-HAADF images. The density is calculated in 10 x1 nm$^2$ areas to average it out.*

Monolayer VMS and MS were grown using a home-built CVD method, as explained in the experimental section and supplementary information,Figure S1a and b. $V_2O_5$ and NaCl powders were added with the $MoO_3$ powder precursor for V doping process. The $V_2O_5$:$MoO_3$ ratio was set to be 1:1, 1:2, and 1:3 for achieving three different doping concentrations of V in VMS and named as VMS-L (low), VMS-M (medium), and VMS-H (high) respectively. Figure 1a shows the optical micrograph of VMS showing a triangular monolayer crystal similar to pristine MS of similar size. A photograph of MS is given in the supporting information, Figure S1c. Different extent of V doping in VMS does not have much effect on the morphology of the crystals.

Figure 1b shows the micro-Raman spectra of VMS and MS. MS shows two typical vibration modes namely $E_{2g}$ (in-plane vibration, sum of degenerate transverse optical (TO) and longitudinal optical (LO) branch) and $A_{1g}$ (out of plane vibration, out of plane optical (ZO)) modes at 384 cm$^{-1}$ and 401 cm$^{-1}$, respectively.[39–41] A separation of ~17 cm$^{-1}$ among $E_{2g}$ and $A_{1g}$ modes confirms the presence of monolayer MS.[39] In the case of VMS, the separation between $E_{2g}$ and $A_{1g}$ peaks enhances from 17 to 26 cm$^{-1}$ depending on the doping concentration. This enhancement resembles the phonon dispersion of vibration modes away from the gamma point (q=0): ZO mode in an upward direction and LO and TO modes in a downward direction (see phonon dispersion of $MoS_2$ in supplementary Figure S2).[40] The $E_{2g}$ and $A_{1g}$ peaks also become broader with increasing doping concentrations. The enhanced full-width at half maximum (FWHM) of both $E_{2g}$ and $A_{1g}$ of VMS with doping is clearly seen in Figure S2 (FWHM of $E_{2g}$ and $A_{1g}$ changed from 4 cm$^{-1}$ and 6 cm$^{-1}$ to ~10 cm$^{-1}$ and ~12 cm$^{-1}$). The broadness of the peaks may represent the presence of localized phonon modes.[41] Other than this broadening, a set of new peaks start emerging in the range 100-260 cm$^{-1}$. These peaks are earlier reported for defect site-related peaks.[40] The vanadium doped sites are also acting as defects for the $MoS_2$ lattice and thus causing these peaks to emerge. To understand them better, Raman signatures of both the regions are deconvoluted: these regions are 100 to 260 cm$^{-1}$ and 350 to 420 cm$^{-1}$ (see inset of Figure 1c). The peaks in the proximity of first-order $E_{2g}$ (at 377 cm$^{-1}$) and $A_{1g}$ (at 411 cm$^{-1}$)

Raman peaks are LO and ZO branches at the M point, respectively.[40] While the four peaks fitted in the 100-260 cm$^{-1}$ region are TA (157 cm$^{-1}$, transverse acoustic) at M, ZA (187 cm$^{-1}$, out of plane acoustic) at M, TA (215 cm$^{-1}$) at K, and LA (227cm$^{-1}$, longitudinal acoustic) at M modes (see phonon dispersion of MoS$_2$ in supplementary Figure S2).[40] LA mode can particularly be used to calculate the average distance ($L_D$) between defect sites in MoS$_2$ (which is V atoms in our case) with the relation:[40]

$$L_D^2 = \frac{I(A_{1g})}{I(LA)} C(A_{1g}) \ ,$$

where *I(A$_{1g}$)* and *I(LA)* are the intensity of A$_{1g}$ mode and LA(M) mode. *C(A$_{1g}$)* is the proportionality constant which is 0.59 nm$^2$ for A$_{1g}$ mode.[40] $L_D$ comes out to be 8.7, 2.6, 1.1, and 0.9 for MS, VMS-L, VMS-M, and VMS-H, respectively (see Figure 1c). From the calculated values of $L_D$, the density of V atoms can be calculated as 1.47 x 10$^{13}$ atoms/cm$^2$, 8.26 x 10$^{13}$ atoms/cm$^2$, and 15.3 x 10$^{13}$ atoms/cm$^2$ respectively for VMS-L, VMS-M, and VMS-H. This gives the value of V doping to be ~1%, ~5% and ~9 % for VMS-L, VMS-M, and VMS-H respectively. These dopant amounts are further confirmed with scanning transmission electron microscope (STEM) and X-ray photoelectron spectroscopy (XPS) based analyses.

The effect of V doping on the chemical bonding of MoS$_2$ is further understood by comparing the XPS of MS with VMS. The survey spectrum of VMS-M shows the presence of V, Mo, and S in VMS, as shown in Figure S3. Figure 1c shows the high-resolution XPS spectra of Mo 3d, S 2p, and V 2p. Mo doublet is observed at 229.4 (3d$_{5/2}$) and 232.7 (3d$_{3/2}$) eV (Figure 1d).[41,42] Apart from this, a small Mo$^{6+}$ peak is also observed at 235.9 eV indicating the presence of some MoO$_3$ particles (ingrown nucleation sites) in the sample.[43] The positions of S 2p doublet are at 162.3 (2p$_{3/2}$) and 163.6 (2p$_{1/2}$) eV (Figure 1e).[42] The V 2p peak is also observed in the XPS. The 2p doublet of V are at 517.2 (2p$_{3/2}$) and 524.3 (2p$_{1/2}$) eV (Figure 1f).[44] The peak positions of Mo and S are observed to be shifted towards lower binding energies (see Figure S3). This shows that the V doping is lowering the Fermi level of MS towards the valence band causing p doping in MS.[44]

V doping is further confirmed using STEM-high angle annular dark field (STEM-HAADF) imaging of VSM-M. Figure 1g display the atomic resolution image of VMS showing the lower intensity points in Mo atomic columns. These lower intensity points are reported to be the presence of V atoms in the MS lattice.[33,45] The distribution of V is found to be random and does not favor clustering.[33] Figure 1h shows the line profiles corresponding to three different locations in Figure 1h. A clear and sudden decrease in the intensity at certain locations confirms the presence of lower atomic number V atoms as the atomic intensities in STEM are dependent on the atomic number variations.[33] The number density of V is calculated manually using three different independent STEM images (see Figure 1i). The value comes out to be ~0.5 atom/nm$^2$ which corresponds to ~4% atomic doping of V. which matches well with the XPS and Raman analyses. All these spectroscopic and imaging techniques confirm the in-plane V doping in the MS lattice leading to the formation of VMS.

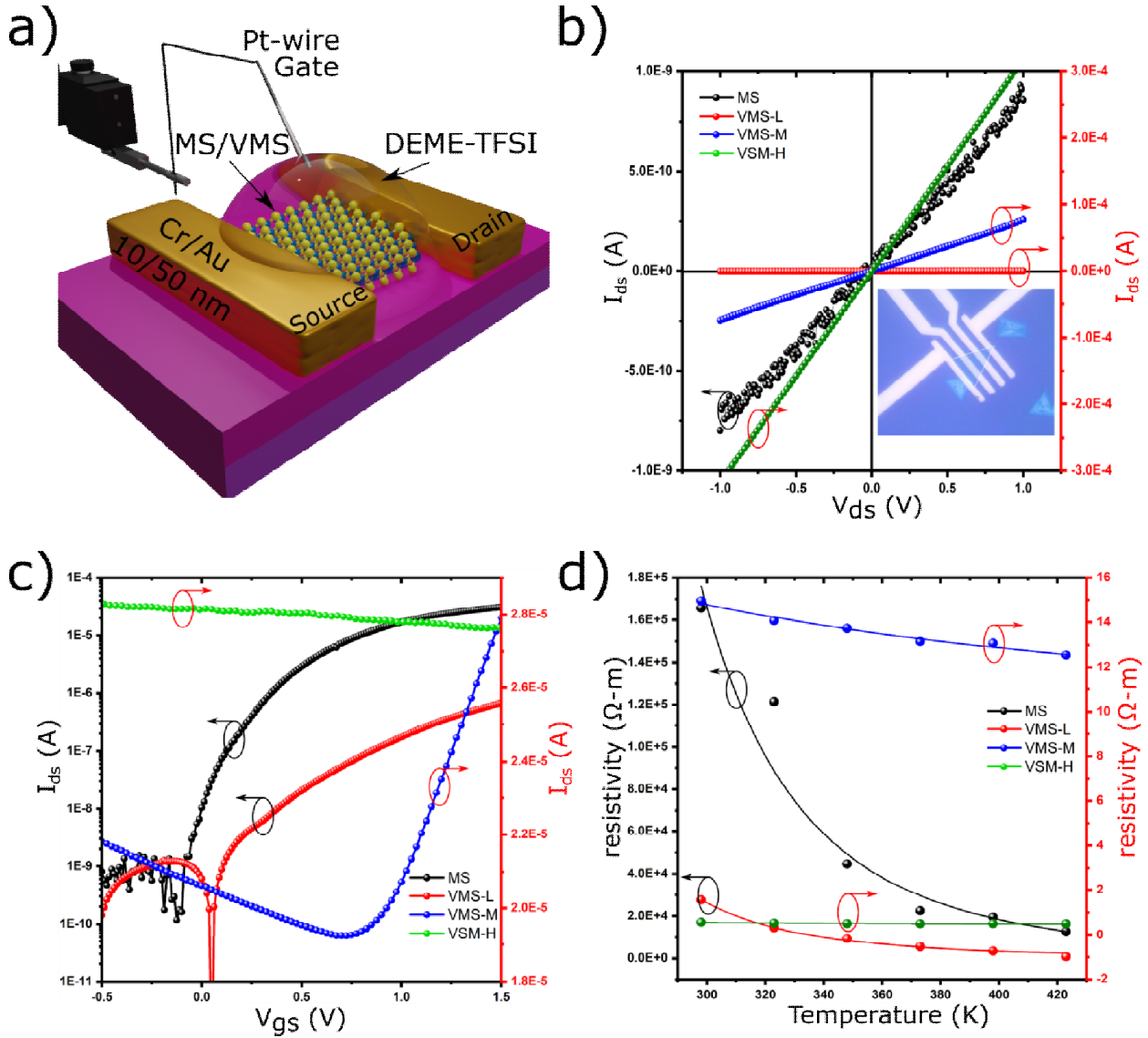

*Figure 2: a) Schematic of MS/VMS based field effect transistor (FET) with DEME-TFSI ionic liquid as (top) gating dielectric material. b) Room temperature four-probe current-voltage (I-V) characteristics curves of MS as well as VMS of three different V concentrations. The y axis*

*(current) for MS and VMS-L/M/H are on the left side with black color and right side with red color, respectively. c) Top gated FET characteristics: $I_{ds}$ vs $V_{gs}$ curves of MS and VMS at $V_{ds}$ = 0.5V. VMS-H is multiplied by 0.5 to put in the same graph for comparison. The y axis (current) is in log scale for MS and VMS-L on the left side with black color and for VMS-M/H on right side with red color, respectively. d) resistivity vs temperature curve for Ms and VMS. The y axis (current) for MS and VMS-L are on the left side with black color and for VMS-M/H on right side with red color, respectively.*

V doping can reduce the Fermi level of MS towards the valence band as suggested by XPS (discussed earlier).[33,44] The same was confirmed with the transport measurements as well. Figure 2a shows the schematic of a device used for the transport measurements of MS, VMS-L, VMS-M, and VMS-H. Figure 2b shows the four-probe current-voltage(*I-V*) characteristics of MS and VMS-L/M/H. The $I_{ds}$ current for MS is in the nA range (drain-source voltage) range from -1V to 1 V at 0 V $V_{gs}$ (gate-source voltage). While in the case of VMS, $I_{ds}$ valuesare found to be in the order of ~$10^{-8}$, $10^{-5}$, and $10^{-4}$ A for VMS-L, VMS-M, and VMS-H, respectively. These very different $I_{ds}$ current ranges can be understood by investigating Ioffe–Regel criterion.[46,47] According to this criterion, $k_F.l_e$>>1 for metals, $k_F.l_e$~1 for semiconductor and $k_F.l_e$<<1 for insulator where $k_F = \sqrt{2\pi n_{2D}}$ is the Fermi wave vector and $n_{2D}$ is the carrier density.[46] $l_e = \frac{hk_F\sigma}{2\pi n_{2D}e^2}$ is the mean free path of electrons and σ is the conductivity. For our four probe devices at $V_{gs}$ = 0V, MS and VMS-L have $k_F.l_e$~ 0.014 and 0.7 respectively, which indicates their insulating nature. While as it is 1.78x$10^3$ and 4.2x$10^4$ for VMS-M and VMS-H respectively,indicating their metallic nature compared to MS and VMS-L. This confirms that high V doping can cause insulator to metal transition in MS.

The effect of V doping is further understood by top-gated field effect transistor (FET) measurements. The measurements are performed using ionic liquid Diethylmethyl (2-methoxyethyl) ammonium bis (trifluoromethylsulfonyl) imide (DEME-TFSI) as dielectric material. The transfer characteristics of MS, VMS-L, VMS-M and VMS-H are shown in Figure 2c. The MS device shows a typical n-type behaviour in the *$I_{ds}$ vs $V_{gs}$* graph with threshold voltage ($V_{th}$) at -0.1V. On the other hand, VMS-L showed $V_{th}$ at 0.06V. The VMS-M shows ambipolar behaviour with $V_{th}$ at 0.8V. The VMS-H being metallic, shows almost independent nature with respect to gating. The positive shift of $V_{th}$ in VMS-L and VMS-M with respect to MS shows p type doping.[17,48] The charge concentration, $n_{2D}$ for MS, VMS-L and VMS-M can be calculated using parallel plate capacitor model by formula:[46,48]

$$n_{2D} = \frac{C_{ox}\Delta V_{gs}}{e},  \quad\quad\quad .1.$$

where $C_{ox}$ is the parallel plate capacitance of dielectric, $\Delta V_{gs} = V_{gs} - V_{th}$($V_{gs}$ is any voltage in On state) and e is the elementary charge, 1.602 x $10^{-19}$ C. At $V_{gs}$ =1.5 V, $n_{2D}$ of electrons comes out to be $7.2 \times 10^{13}$, $6.48 \times 10^{13}$, $3.15 \times 10^{13}$ cm$^{-2}$ for MS, VMS-L and VMS-M respectively. This significant and constant decrease in electron density with V doping clearly shows the V atoms are acting as intrinsic p dopants for MS. The same can be understood by PL fitting in supplementary Figure S4 where A- trion contribution significantly reduced in the total PL of VMS-L compared to MS.

Effect of V doping on the work function of MS can be seen by calculating it using formula for change in work function ($\Delta \Phi$):[48]

$$\Delta \Phi = \Phi_{VMS} - \Phi_{MS} = ln(\frac{n_{2D,VMS}}{n_{2D,Ms}}), \quad\quad\quad .2.$$

where $\Phi_{VMS}$ and $\Phi_{MS}$ are work functions of VMS and MS respectively. $n_{2D}$ is the carrier density extracted from FET measurements of MS and VMS as discussed above. The calculated $\Delta \Phi$ for VMS-L and VMS-M is ~105 meV and ~826 meV respectively. The calculation shows that the fermi level shifts significantly with V doping. This observation matches well with our DFT based theoretical values where $\Delta \Phi$ changes ~720 meV for ~4% V doping (discussed in experimental methods in details).

The slope of the linear region of the transfer curve was then used to calculate the mobility of electrons and holes with the formula:[46,49]

$$\mu_{e,h} = \frac{L}{WV_{ds}C}\left(\frac{dI_{ds}}{dV_{gs}}\right)_{e,h} \quad\quad\quad .3.$$

where, µ is the mobility for an electron or hole, L and W are the length and width of the channel device respectively. C is the gate capacitance per unit area (7.2 µFcm$^{-2}$) and $\left(\frac{dI_{ds}}{dV_{gs}}\right)$ is the slope of the transfer curve.[50] The $\mu_e$ for MS is calculated as ~7.8 cm$^2$V$^{-1}$s$^{-1}$. In case of VMS-M, $\mu_e$ and $\mu_h$ are found to be 26 cm$^2$V$^{-1}$s$^{-1}$ and 17 cm$^2$V$^{-1}$s$^{-1}$, respectively.

The temperature dependent resistivity was also calculated for MS and VMS L/M/H in high temperature range (300-473 K). All the systems are found to have thermally activated transport.

Thus, the temperature dependence of resistivity of MS, VMS-L/M/H was fitted with the Arrhenius equation:[46]

$$\rho = \rho_o e^{\frac{E_a}{K_b T}}, \qquad .4.$$

where $\rho$ is the resistivity, $K_b$ is the Boltzmann constant and $E_a$ is the activation energy for thermally activated transport. The $E_a$ comes out to be 214, 196, 21, and 10 meV for MS, VMS-L, VMS-M and VMS-H, respectively. The activation energies for VMS-M and VMS-H are less than the room temperature energy (~25.7 meV) which makes them metallic at room temperature.

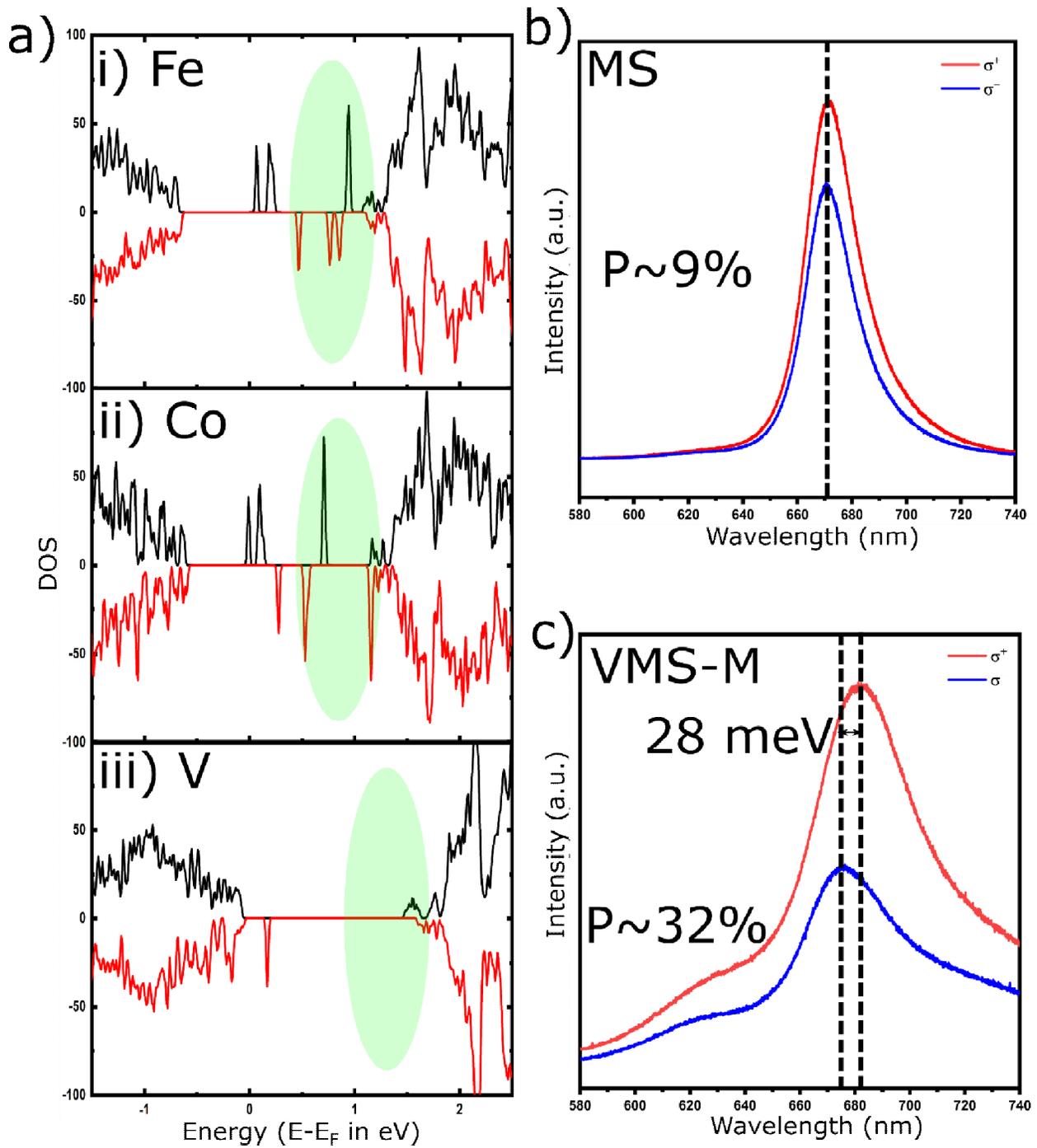

*Figure 3: Density of states (DOS) of MS with i) Fe ii) Co and iii) V doping. The doping concentration is 4% in all cases. Black and red color are for spin up and spin down states respectively. A light green highlighted circle is to show sharp DOS in MS in presence of magnetic doping. Helicity dependent PL from b) MS and c) VMS-M.*

The transport measurements show that the V doping not only tune the electrical properties of MS but at the same time it can change its behaviour from n-type to ambipolar to metallic material. One of the main and interesting aspects of V doping is its ability to induce magnetism in MS which has already been mentioned in many theoretical reports.[38] But in most magnetic doping cases, a high density of state (DOS) band arises near the conduction band.[17,38] Photoexcited electrons hop into these dense state bands which results in PL quenching.[17] These bands usually provide non-radiative pathways to the photoexcited electrons. This situation is sometimes called the exclusion rule of coexistence of PL and ferromagnetism in TMDs.[17] The same can be understood by Figure 3a where DOS is shown for Fe, Co and V doped $MoS_2$. The atomic percentage of doping is 4% in all the three cases. Figure 3a clearly shows spin polarization in all three cases. However, Fe and Co show some sharp peaks near the conduction band (Figure 3 i and ii, light green highlighted region). These peaks can actually reduce the PL of MS and thus the exclusion rule works here.[17] But in case of VMS, these sharp peaks are almost absent and thus V doping can maintain the PL of MS. Secondly, V alone has an electronic configuration of *$3d^4 4s^1$* which has one less electron compared to Mo (*$4d^5 5s^1$*).[38] This can induce 1 $\mu_B$ magnetic moment per V atom into VMS.[38] In case of V doping, the induced spin on six nearest neighbor Mo atoms is parallel to that of V which results in an enhanced magnetic moment in VMS systems.[38] The discussion suggests a ferromagnetism with V doping in $MoS_2$. Existence of spin polarisation at the Fermi level indicates the presence of ferromagnetism in VMS, as seen in V doped $WS_2$ and $WSe_2$.[34,37] Thus, a possible breaking of the exclusion rule of PL and ferromagnetism in case of VMS is expected, making it an interesting material to work on. In our experimental results, we find a significant PL in case of both VMS-L and VMS-M. Helicity dependent PL measurements were performed using 532 nm laser, quarter waveplate and a half waveplate. The polarization in PL is calculated using the formula:[37,51]

$$P = \frac{I_+ - I_-}{I_+ + I_-}, \qquad .5.$$

where $I^+$ and $I^-$ are the intensity of the PL with excitation $\sigma^+$ (right circular, say) and $\sigma^-$ (left circular) helicity light. The P comes out to be ~9% for pristine MS (Figure 3b), while in the case of VMS-M, it comes out to be ~32% (Figure 3c). Even if we consider the 9% difference as due to additional path length in the set up due to the quarter wave plate, the huge difference in polarization of VMS can be assumed due to the valley polarization. Furthermore, VMS also showed a shift of ~28meV with $\sigma^+$ and $\sigma^-$ helicity light excitation. This shift signifies the breaking of degeneracy (time reversal symmetry breaking) in K and K' valleys with different energies of valence band and conduction bands. This shows that the VMS not only has a highervalley polarization but can also show significant valley splitting even at room temperatures.

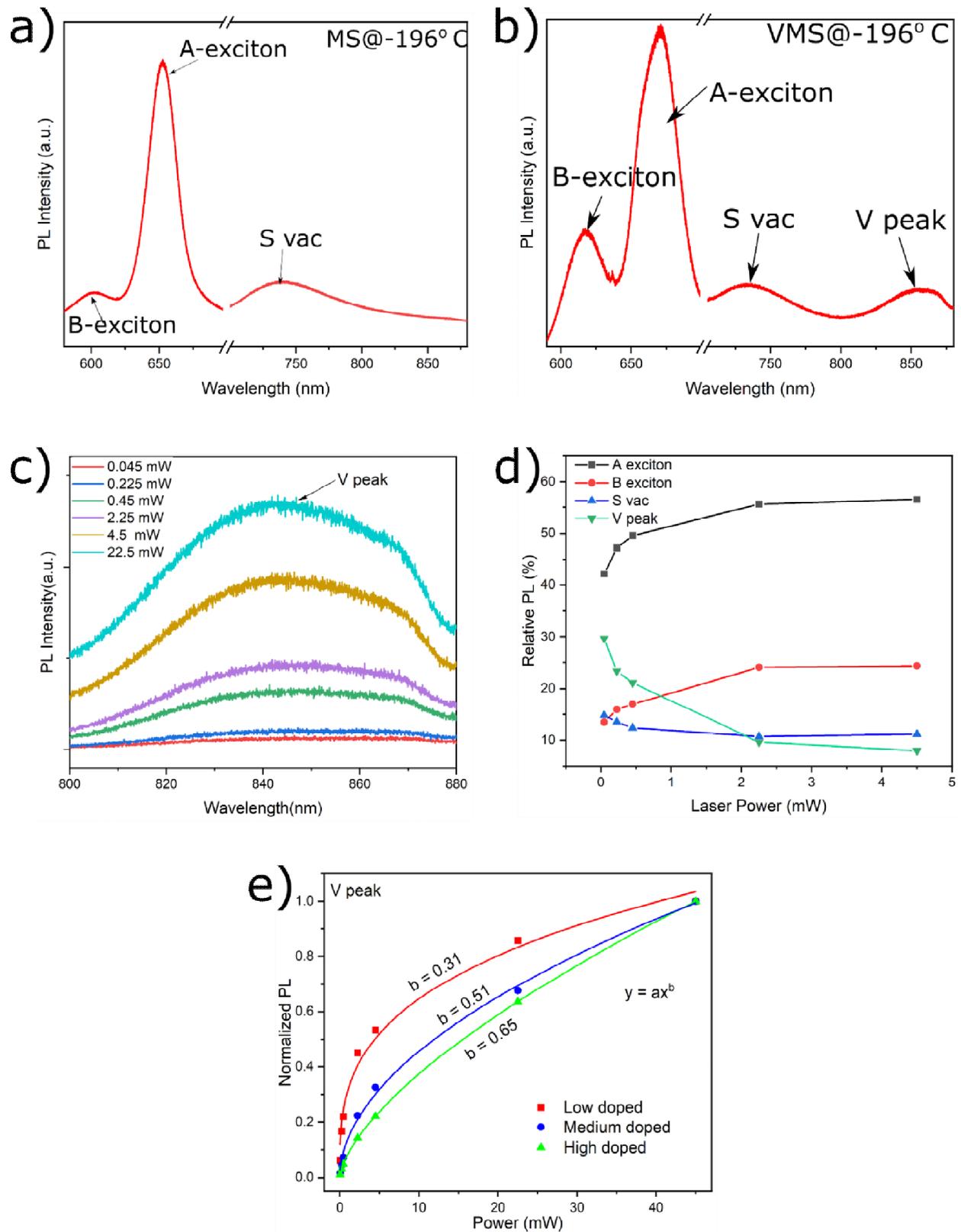

*Figure 4: Photoluminescence at -196 $^0$C from a) MS b) VMS-M. c) Power-dependent PL*

*measurement. d) Laser power vs Relative PL(%) plot. e) Normalized PL vs Power plot of V peak for three different doping concentrations.*

Temperature dependent PL measurements are also carried out on both MS and VMS-M samples. The details of measurements can be found in the experimental section and supplementary Figure S5. Figures4a and b are showing the PL spectra at -196 $^{o}$C for pristine MS and VMS-M samples. The MS PL shows two well-known A and B excitonic peaks arising from the valence band splitting due to high spin-orbit coupling in MS.[41,52–54] A new peak can also be observed at ~720 nm which can be attributed as a defect peak.[53] The most common defects that can be observed in as grown MS are single chalcogen vacancies ($S_v$), double chalcogen vacancies ($S_{2v}$), and metal vacancies ($M_v$).[55] The formation energies of these respective defects are calculated in our previous work.[41] The formation energy of $M_v$ is ~ 5 eV while for $S_v$ and $S_{2v}$ the formation energy is around 2 eV and 3 eV, respectively.[41] Thus, $M_v$ is rare to be observed in low-temperature PL of MS.[53] Earlier low temperature PL reports have already shown sulfur vacancy related PL peaks at ~720 nm in MS.[41,53] Thus, the peak at 720 nm can be attributed to sulfur vacancy defects and is named as S vac in the upcoming discussion. The low-temperature PL of VMS not only shows A exciton, B exciton, and S vac, but also shows an extra peak around 850 nm. This new peak is observed in all low doped, medium doped, and high doped VMS (see supplementary Figure S6) but not in pristine MS. We called this new peak as the V peak in an upcoming discussion. The evolution of $S_{vac}$ and V peak with temperature is shown in Figure S5. Both peaks are found to be Arrhenius in nature, which activates below a certain temperature. The S peak starts to appear at ~ -80 $^{o}$C while the V peak starts to appear at ~-40 $^{o}$C as shown in Figure S5.

The power-dependent PL for VMS is carried out where V peak increases monotonically with laser power as shown in Figure 4c.[41] To understand its behavior better, the relative integrated PL (RIPL) contribution of each peak is plotted with respect to the power in Figure 4d. The contribution of RIPL from A and B excitonic peaks increases continuously with laser power while for S and V peaks,it decreases as the laser power increases (see Figure 4c).[53] The behaviors of the S and V peaks are observed to be the same with power. As the defect sites are constant in number and don't increase with laser power in the given power range, the RIPL is expected to decrease because of Pauli blocking.[53] In the case of VMS, the extra states may arise in the band gap of MS due to V doping. These states are also constant in number in a crystal and depend only on the doping concentration of V. Thus, showing a decrease in RIPL of V peak as the states get filled easily with power.

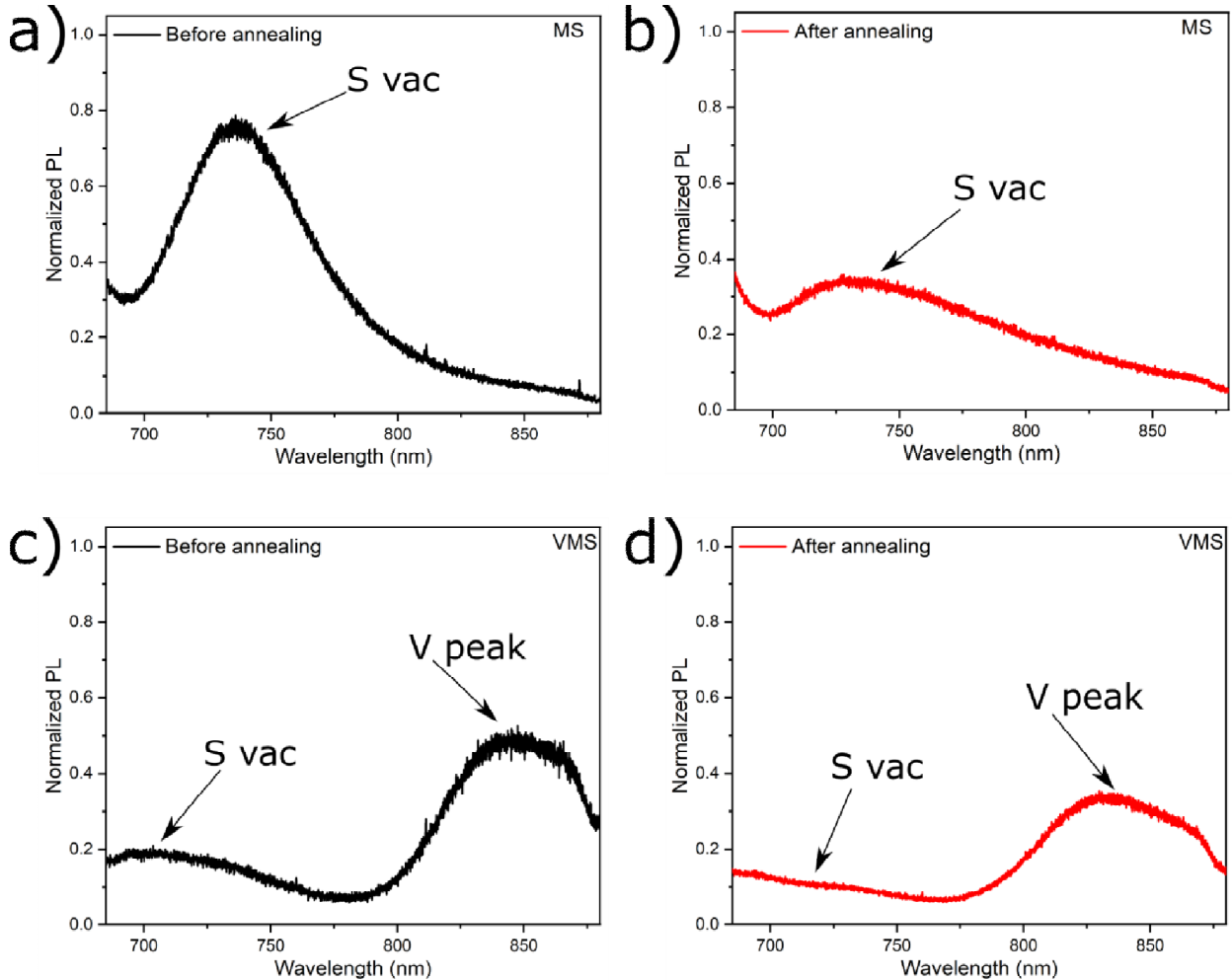

*Figure 5: Normalized PL measurement at -196 $^oC$ for a) MS before annealing b) after annealing c) VMS-M before annealing d) VMS after annealing. All figures are normalized with respect to their A exciton.*

This claim is further confirmed with the power-dependent PL measurements of V peak for low, medium, and high doped VMS. The PL intensity increases linearly for A and B excitonic peaks as shown in Figure S7 because of the availability of empty states in the conduction band. However, in the case of VMS, states belonging to V peak are expected to be proportional to V concentration. Thus, the V peak intensity is sub-linear with respect to power (Figure 4e). The proportionality constants are found to be ~ 0.31, 0.51, and 0.65 for low, medium, and high doped VMS, respectively. This clearly indicates that the mid-gap states increase with the increase in V concentration. Secondly, the increase in V doping decreases the overall PL intensity as shown in Figure S6.

To further confirm that the origin of the V peak is different from the $S_v$ or $S_{2v}$ vacancies, sulfur annealing carried out in both MS and VMS at 200 $^0$C for 1 hour and the PL spectra after annealing are shown in Figure 5. Note that all the graphs in Figure 4 are normalized with respect to their A exciton peak. In the case of MS, the S peak decreases significantly after sulfur annealing as shown in Figures 5a and b.[56] The S vapors refill the S vacancies and thus reduce the S peak in MS. In the case of VMS, the S annealing decreases the S peak as the S vacancies decrease[56] but the V peak intensity remains the same as shown in Figures 5c and d. This confirms that the origin of the V peak is different from the S peak and is coming from the V doping only.

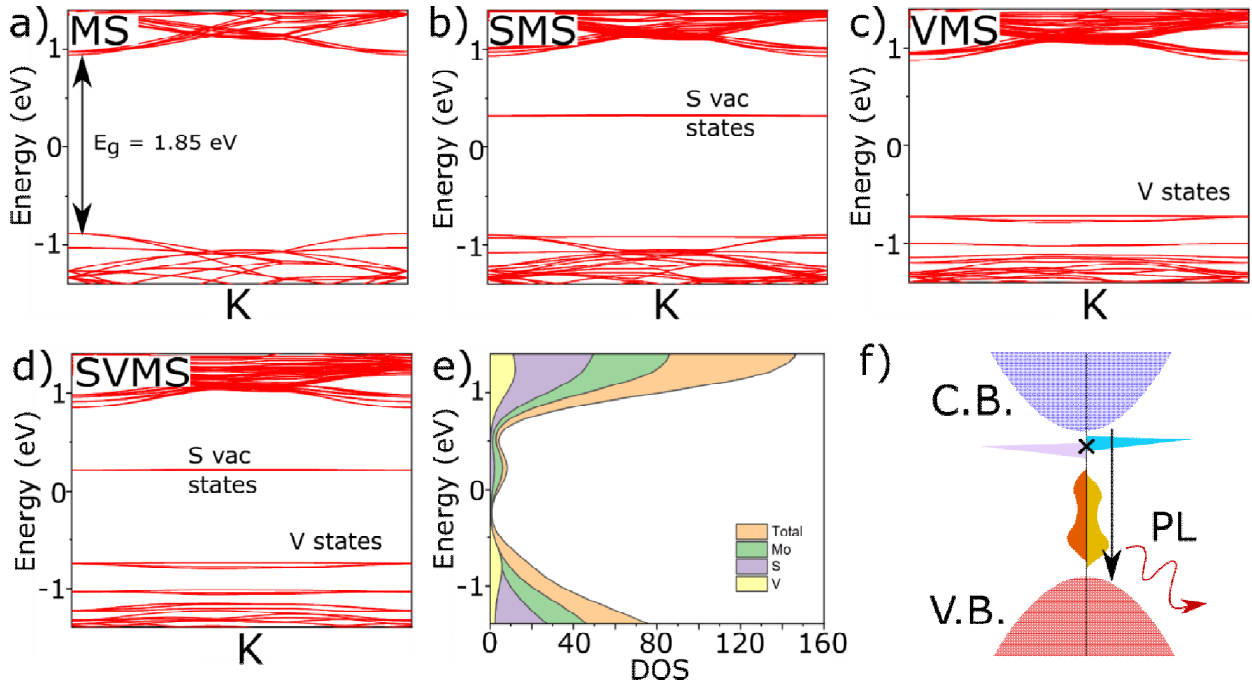

*Figure 6: Energy band diagram plot using DFT for a) MS b) MSlattice with a S vacancy, SMS c) V doped MS, VMS d) V doped MS with a S vacancy, SVMS. e) Partial density of state for SVMS showing orbital contribution from different atoms. f) Schematic of VMS band structure which*

*doesn't have dense states near conduction band and hence favors the radiative recombination of photoexcited electrons.*

A DFT based study on stabilized structures of monolayer $MoS_2$ (MS), MSlattice with a S vacancy (SMS), V doped MS(~12% atomic doping) (VMS), and V doped MS with a S vacancy (SVMS) is carried out (details of DFT study are given in the experimental section and supporting information). The respective unit cells can be seen in Figure S8. MS shows a direct band gap of ~ 1.85 eV at high symmetry K point in momentum space, which is in tune with the experimental PL values observed here and also with the reported literature.[13,57,58] In SMS, the band gap remains the same but some extra states (deep level) are emerging in the band gap.[24] These states are arising due to the S vacancy in the system.[24,41] These states are closer to the conduction band and can be the reason for S peak in the low-temperature PL.[41] The VMS structure has three V atoms in the MS lattice. All the V atoms are placed close to each other to minimize the formation energy of the system. The band structure of VMS also shows some mid-gap states in the band structure. The states are closer to the valence band.[38] These states can be a reason for V peak in the PL at lower energies than A-excitonic emissions. In the case of SVMS, the band structure has two types of mid-gap states. One closer to the conduction band is from the S vacancy while the ones closer to the valence band are from the V atoms. The same is confirmed with a partial density of state calculations in Figure 6e. The S vac states (near conduction band) are majorly from the Mo atoms. In absence of S atoms, the vacant d orbitals of Mo are contributing to the mid-gap states. While the V states (near valence band) are majorly coming from the V atoms with small contributions from the Mo and S states. The presence of these two types of mid-gap states can be a very possible reason for two peaks i.e., S vac and V peak in VMS crystals at low temperatures.

## 3. **Conclusions**:

Here, we showed the possibilities of the existence of photoluminescence and spin polarization at room temperature by substitutional vanadium (V) doping of $MoS_2$ monolayers (VMS). It is demonstrated that the variation in the extent of V doping in $MoS_2$can transform it from semiconducting (1%, VSM-L) to metallic (9%, VSM-H) in nature. The ambipolar nature of VMS -Mis shown using ionic liquid assisted top gating-based FET measurements, and the electron and hole mobilities of VMS are calculated as 26 cm$^2$V$^{-1}$s$^{-1}$ and 17 cm$^2$V$^{-1}$s$^{-1}$, respectively. The coexistence of PL, valley polarizations(~32%), and valley splitting (28 meV shift in PL with helicity $\sigma^+$ and $\sigma^-$ of light) have been shown for VMS while valley splitting is not seen in pristine MoS$_2$. A new excitonic peak at ~850 nm has also been observed at low temperature PL measurements and it has been identified due to the V-states present near the valence band. The origin of this PL is further confirmed with DFT based calculations and a detailed band structure analysis. This controlled V-doping along with a detailed spectroscopic, microscopic, and transport measurements establish VMS as an important material for opto-spintronics and establishes the possibilities of engineering TMDs at room temperature for resulting such potential materials.

**Experimental Methods**:

*MS and VMS Synthesis*: VMS synthesis was carried out using CVD assisted method having quartz tube of length ~ 120 cm and inner diameter ~ 5cm. A schematic of the synthesis method is shown in the supporting information, Figure S1. In this method, MoO$_3$ (5 mg) and vanadium oxide powders (5-15 mg) are used as precursors, where they are kept in an alumina boat. The Si/SiO$_2$ (300 nm) substrate was kept on the top of the boat. Another alumina boat is used for sulfur (300 mg), and both these boats were placed inside the CVD quartz tube while keeping the sulfur boat at the left and the precursor boat at the right, as shown in Figure S1. The distance between these two boats was kept ~15 cm. The growth temperature of the sulfur side was 210 °C and the temperature of the precursor side was 710 °C. The temperature ramp time to achieve 710 °C was kept as 25 minutes and the growth time was ~15 minutes with a gas flow rate of ~185 sccm.

In order to ensure the absence of sulfur vacancies in MS and VMS grown *via* the above-mentioned process, further annealing of the samples with sulfur was also carried out. Sulfur annealing was carried out at 180 °C with a nitrogen flow rate of 100 sccm for 60 minutes.

*Transmission electron microscopy measurements*: High-resolution scanning transmission electron microscopy (HRSTEM) analyses have been conducted in a probe-corrected Thermo Fisher Scientific Titan microscope, operated at 120 keV, fitted with a X-FEG® gun and Cs-probe corrector (CESCOR from CEOS GmbH) and a convergence angle of 25 mrad. An Oxford

Instruments Ultim X-MaxN 100TLE detector was used for energy-dispersive X-Ray spectroscopy (EDX) analysis.

*X-ray photoelectron spectroscopy (XPS)*: XPS data was recorded on a Kratos Axis Supra spectrometer equipped with a monochromated Al Kα X-ray source using an analyzer pass energy of 160 eV for survey spectra and 20 eV for the core level spectra. Spectra were recorded by setting the instrument to the hybrid lens mode and the slot mode providing approximately a 700 x 300 $\mu m^2$ analysis area using charge neutralization.

*Low-temperature Raman and PL measurements*: Low-temperature Raman and PL spectroscopic studies were carried out using a low temperature liquid nitrogen setup (Linkham cell HSF-600). The Raman studies (using Renishaw inVia Raman spectrometer) are conducted using 532 nm laser excitation and PL studies using the Raman spectrometer are conducted with 633 nm excitation. The PL measurements are taken from Room temperature to -196$^o$C, where the spectrum at each temperature of both heating and cooling are verified.
In order to conduct polarization dependent PL measurements, the vertically polarized 532 nm laser light is passed through a quarter wave plate and a half waveplate.

*Transport Measurements:* Devices were fabricated using Laser lithography. The metal deposition for the devices was done using thermal evaporator. Finally, the devices are developed in acetone with mild sonication. All measurements were done in ambient environment with Keithley 2450 source meter.

*DFT Studies:* First principle DFT-based calculations were carried out using QUANTUM ESPRESSO 6.8 package with a plane-wave basis set (50-Ry cutoff) and norm-conserving pseudopotentials. The exchange-correlation energy of electrons was approximated with a local density approximation and a parametrized functional of Perdew and Zunger. A vacuum of ~20 Å was added, separating adjacent periodic images along the z-direction. The lattice parameter of monolayer MoS2 unit cell was optimized to 3.1667 Å using Brillouin zone sampling of 12 x 12 x 1. Further, lattice vibrational modes of the unit cell were calculated using uniform grids (nq1, nq2, q3), where n = 4. Further structures of MoS 2 were simulated using a 6x6 periodic supercell with different levels of V doping and S vacancies, where integrations over the Brillouin zone was sampled with uniform $2 \times 2 \times 1$.

**Acknowledgments:**

Authors from TIFR thank the support of the Department of Atomic Energy, 350 Government of India, under Project Identification No. RTI 4007. TNN acknowledges the financial support from DST-SERB, Govt. of India SUPRA scheme (SPR/2020/000220). RS acknowledge Dr. SumitBawari for the discussion regarding DFT calculations. The TEM and XPS studies were performed in the Laboratorio de MicroscopiasAvanzadas (LMA), Universidad de Zaragoza (Spain). We thank G. Antorrena (LMA) for the XPS measurements. R.A. acknowledges support from Spanish MICINN (PID2019-104739GB-100/AEI/10.13039/501100011033), Government of Aragon (projects DGA E13-20R (FEDER, EU)) and from the European Union H2020 programmes "ESTEEM3" (Grant number 823717) and Graphene Flagship (881603).

**TOC**

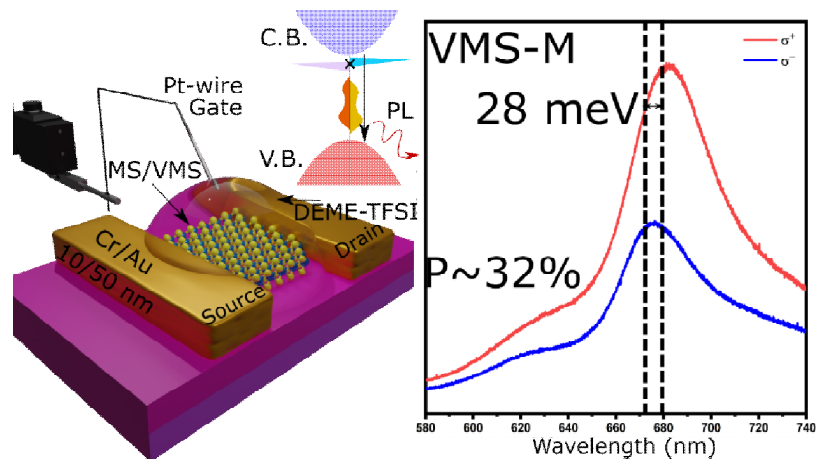

Ambipolar Vanadium (V) doped $MoS_2$ Monolayers are shown for their room temperature valley splitting and spin-polarisation along with emergence of photoluminescence due to V doping, indicating their potential in Opto-Spintronics.

# On the Existence of Photoluminescence and Room-Temperature Spin Polarization in Ambipolar V doped MoS$_2$ Monolayers

Dipak Maity[1], Rahul Sharma[1,*], Krishna Rani Sahoo[1], Ashique Lal[1], Raul Arenal[2,3,4], Tharangattu N. Narayanan[1,*]

[1]Tata Institute of Fundamental Research – Hyderabad
500046, INDIA.

[2]Instituto de Nanociencia y Materiales de Aragon (INMA), Universidad de Zaragoza
50009 Zaragoza, SPAIN.

[3]Laboratorio de Microscopias Avanzadas (LMA), Universidad de Zaragoza, 50018 Zaragoza, SPAIN.

[4]Fundación ARAID, 50018 Zaragoza, SPAIN.

(*corresponding Authors tnn@tifrh.res.in, rahulsharmatifr@gmail.com)

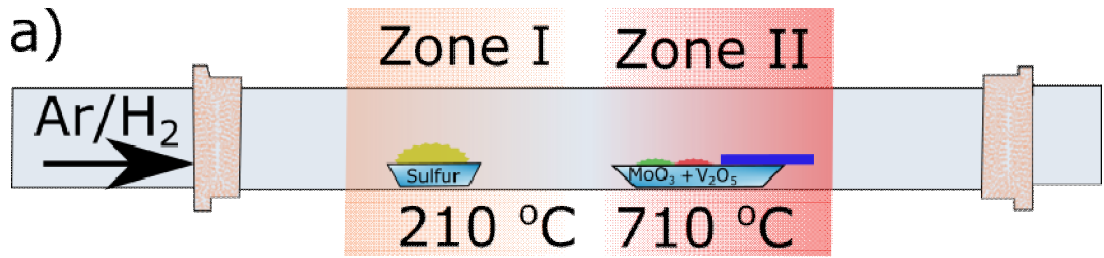

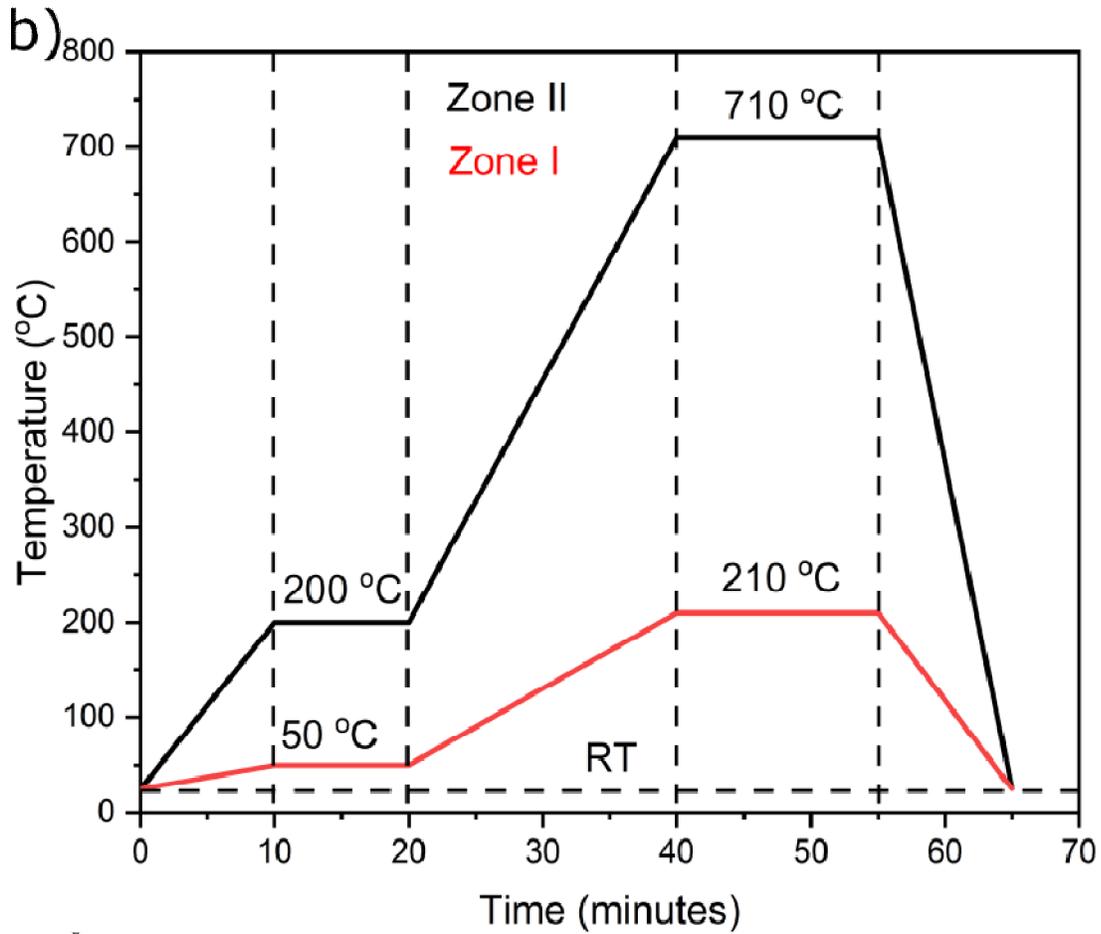

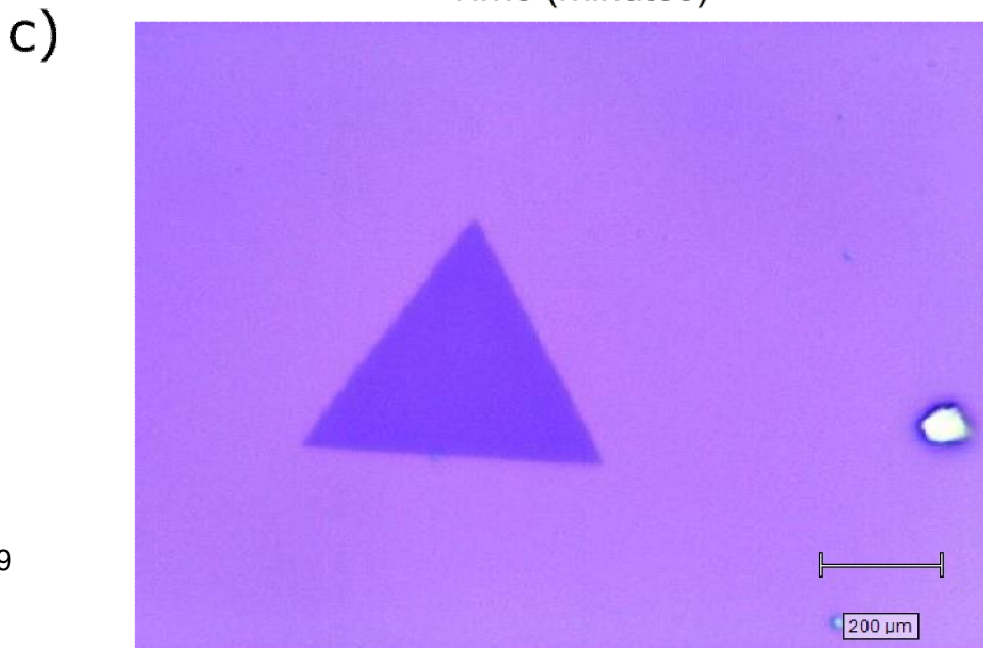

Figure S1: a) CVD set up for V doped $MoS_2$ (VMS) growth. b) Time temperature profile for VMS growth. c) Optical image of monolayer $MoS_2$ (MS).

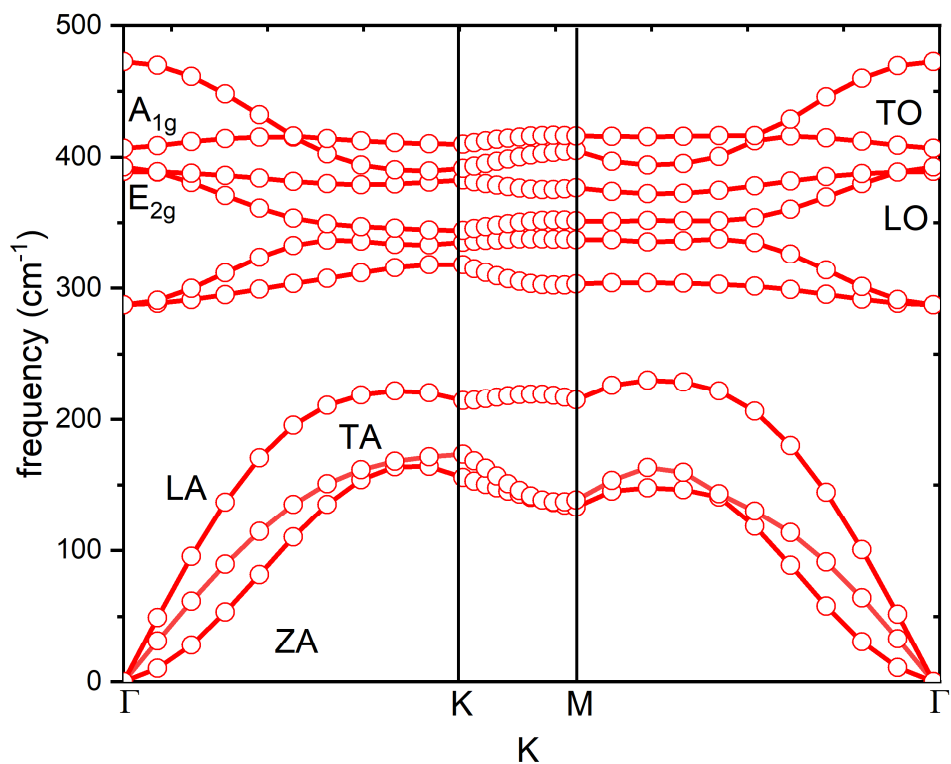

Figure S2: Phonon dispersion curve of $MoS_2$.

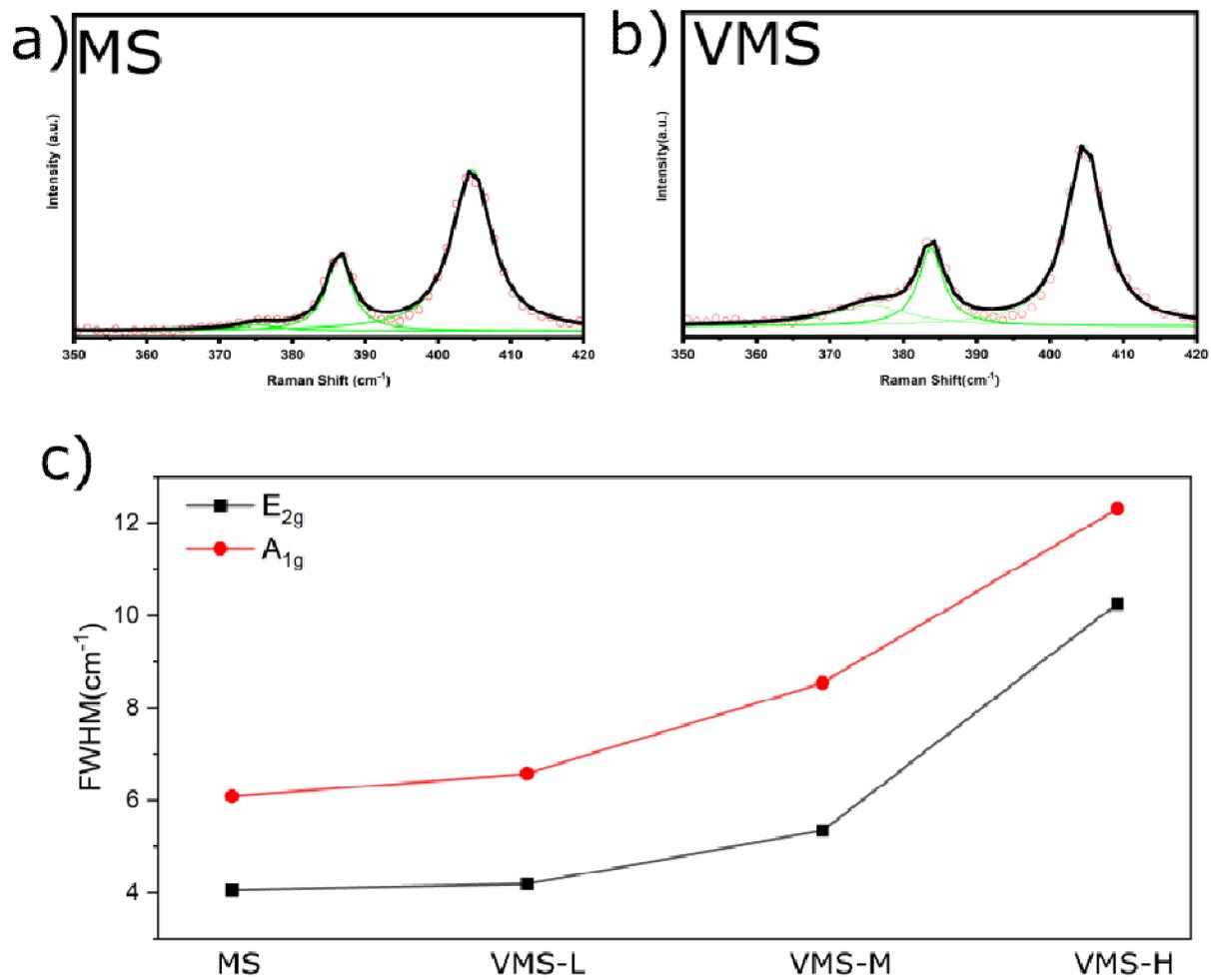

Figure S3: Raman spectra fitting for a) MS b) VMS. c) Full width half maxima (FWHM) of $A_{1g}$ and $E_{2g}$ *vs* V doping.

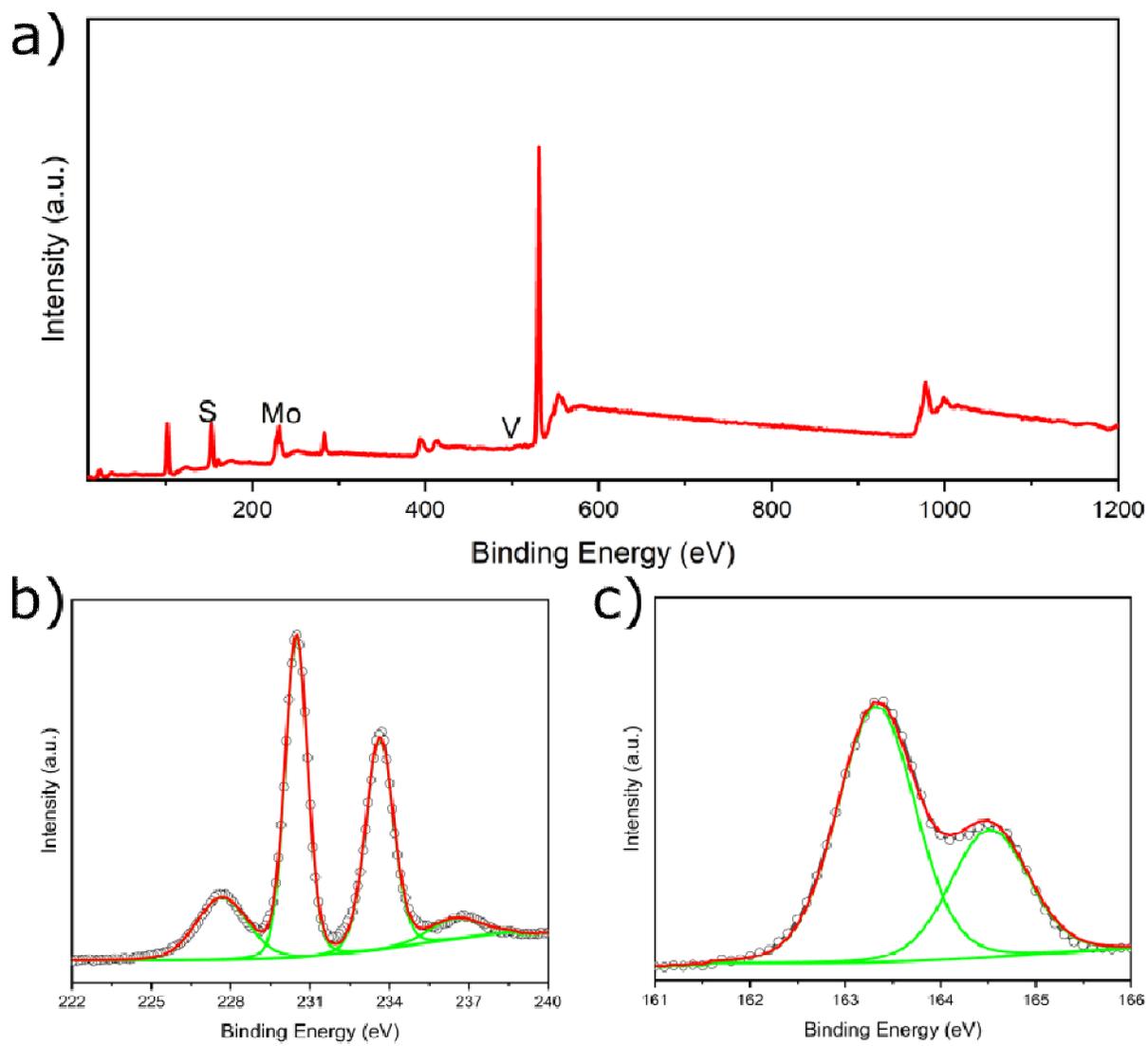

Figure S4: a) Survey spectra of VMS. b) high resolution XPS of Mo 3d. c) high resolution XPS of S 2p.

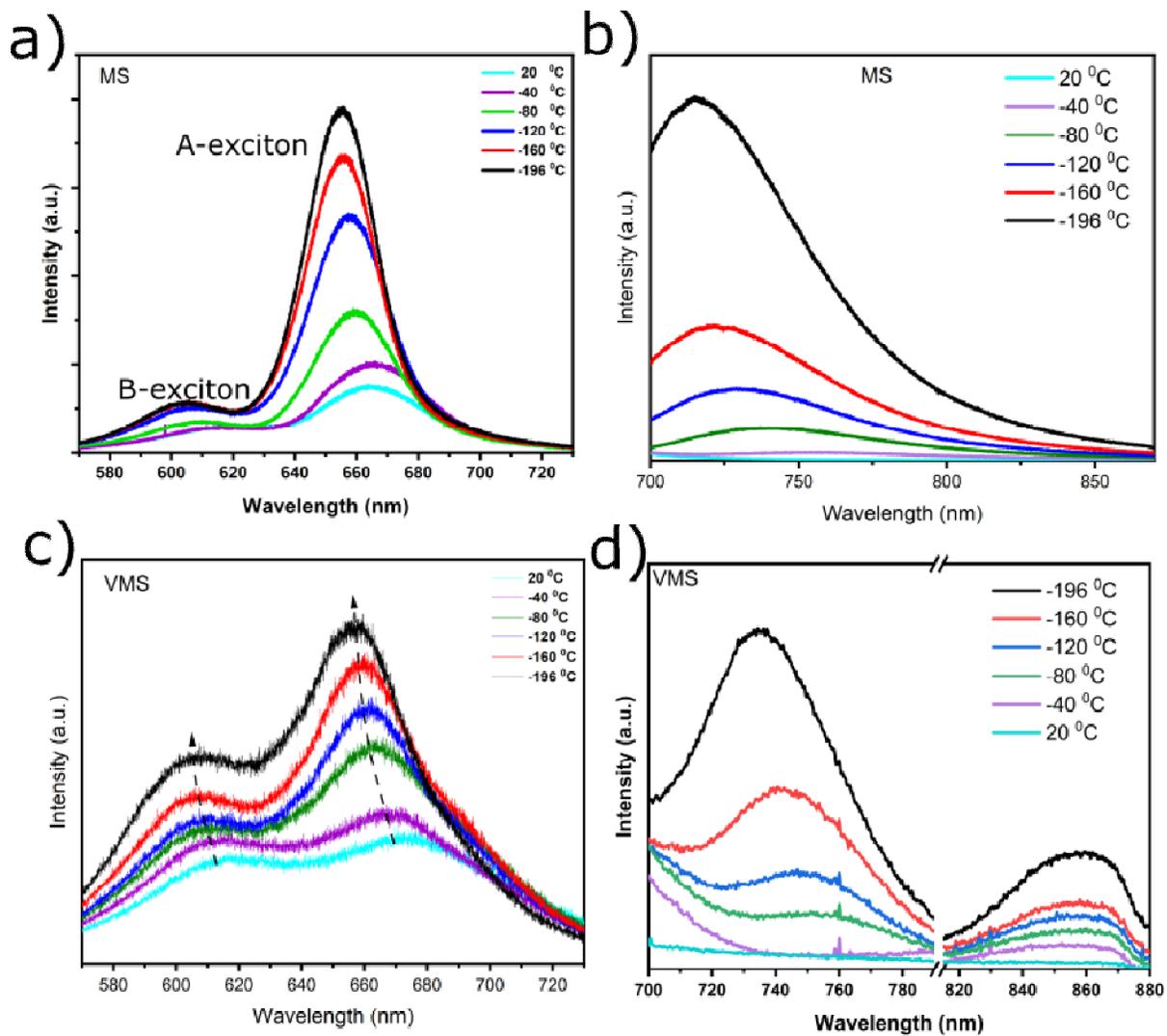

Figure S5: Temperature dependent PL for MS a) 532 nm excitation laser b) 633 nm excitation laser. Temperature dependent PL for VMS c) 532 nm excitation laser d) 633 nm excitation laser.

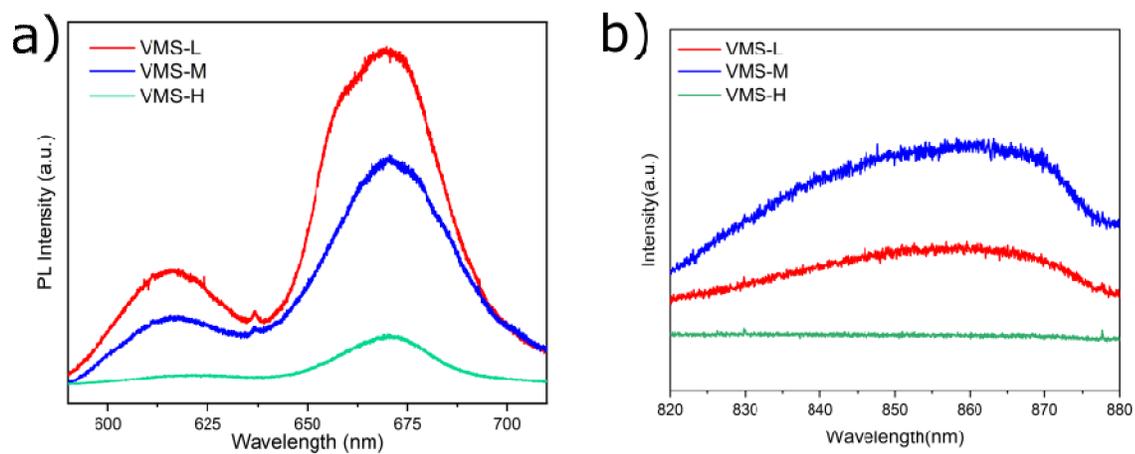

Figure S6: concentration dependent VMS PL @-196°C a) 532 nm excitation laser b) 633 nm excitation laser.

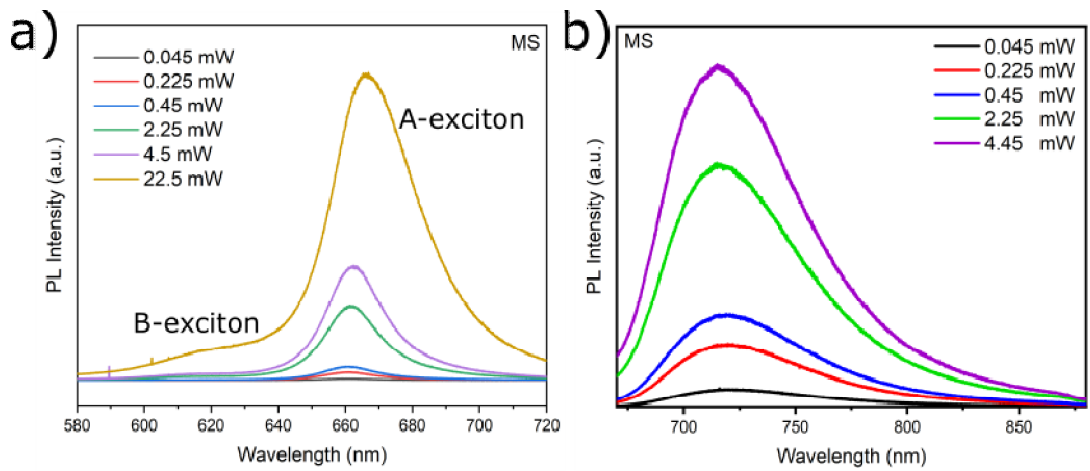

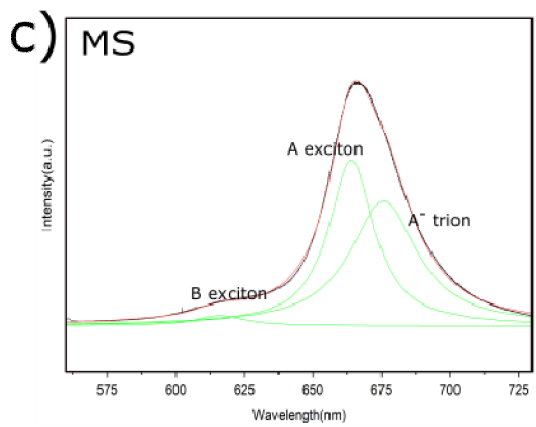
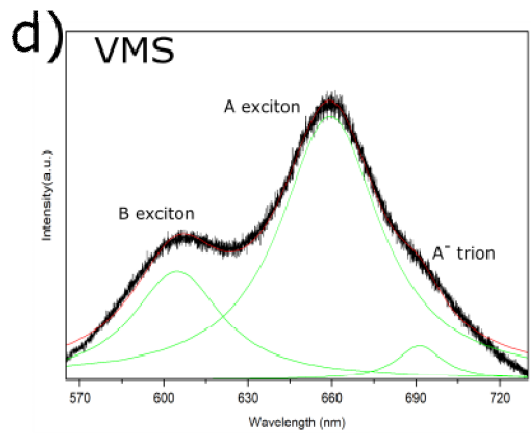

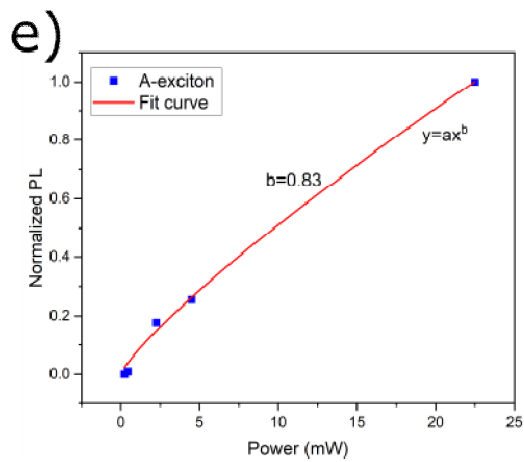
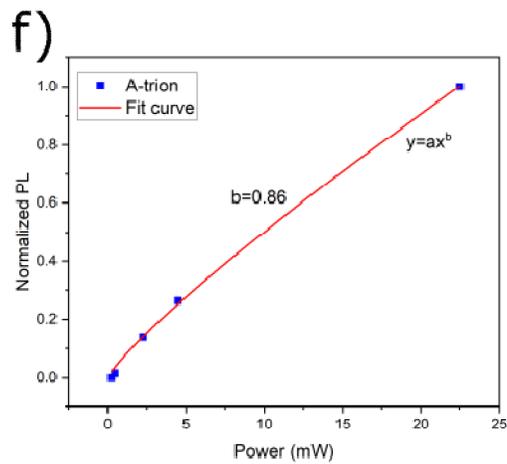

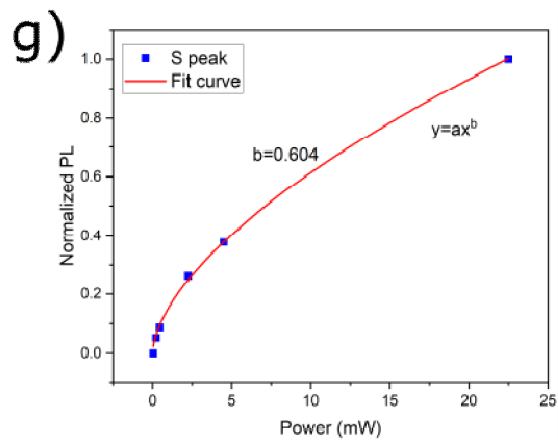

Figure S7:Power dependent PL for MS a) 532 nm excitation laser b) 633 nm excitation laser.c)PL fitting of MS for 532 nm excitation laser (from PL fitting, A trion / A exciton =0.8) .d) ) PL fitting of VMS-M for 532 nm excitation laser (from PL fitting, A trion / A exciton =0.1) . Normalized PL vs Power plot for e)A-exciton f) A-trion g) S peak.

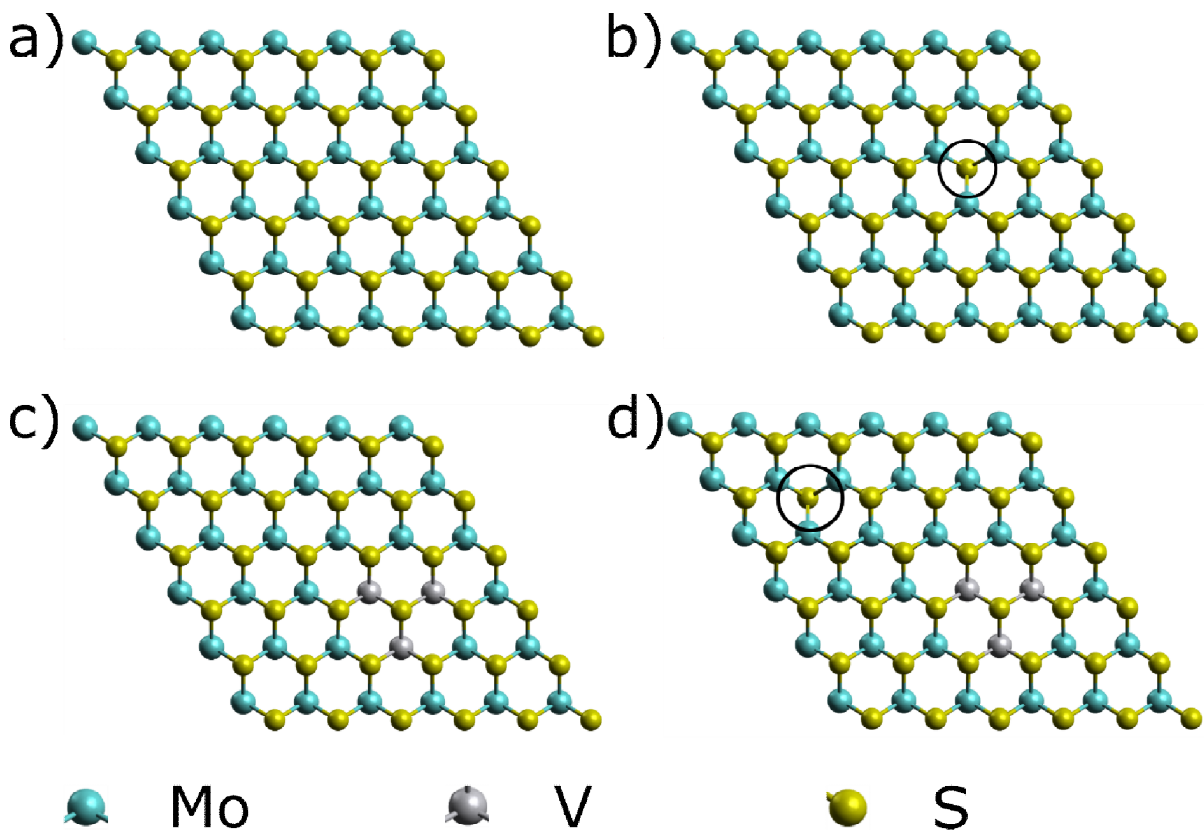

Figure S8: Atomic structure used for DFT calculations: a) MoS2 (MS) b) MS with single sulfur vacancy (SMS). C)  MS with V doping (VMS) d) VMS with single sulfur vacancy.